\documentclass[12pt]{article}
\usepackage{psfig}
\begin{document}
\begin{center}
\textbf{\Large Experimental Investigations of Changes \\ in $\beta$ -Decay rate of
 $^{\rm 60}$Co and $^{\rm137}$Cs.}

\bigskip

\textit{Yu.A. Baurov, A.A. Konradov*, V.F. Kushniruk**, E.A. Kuznetsov**,\\
Yu.V. Ryabov***, A.P. Senkevich***, Yu.G. Sobolev**, S.V. Zadorozsny***.}
\bigskip

\textit{Central Research Institute of Machine Building, \\
141070, Korolyov, Moscow Region, Pionerskaya, 4. Russia.}\\
\textit{* Institute of Biochemical Physics of Russian Academy of Sciences, 117977,
Moscow, Kosygin-Street, 4, Russia.}\\
\textit{** Flyorov Laboratory of Nuclear Reactions, Joined Institute for Nuclear
Research (JINR), 141980, Dubna, Moscow Region, Russia.}\\
\textit{*** Institute for Nuclear Research of Russian Academy of Sciences (INR RAS),
B-312, Moscow, Prospect 60-letiya Oktyabrya, 7a, Russia.}

\end{center}

\abstract{Results of simultaneous measurements of $\beta$-decay rate with the aid of
Ge(Li)-detectors performed at two laboratories 140km apart (INR RAS,
Troitsk, $^{{\rm 60}}$Co, and JINR, Dubna, $^{{\rm 137}}$Cs) during a period from 15.03.2000 till 10.04.2000, are presented.
Regular deviations of the count rate of $\gamma$-quanta following the $\beta$-decay of
$\sim $ 0.7\% (INR RAS,$^{{\rm} {\rm 60}}$Co) and $\sim $ 0.2\% (JINR,
$^{{\rm 137}}$Cs) from the statistical average, are observed.
The analysis of extremum deviations of $\gamma$-quanta count rate shows that the
set of directions of tangents to the Earth's parallels of latitude at the
extremum points of trajectories of motion in the space of each laboratory
clearly forms three separate compact subsets of directions which agree, for
two laboratories, to an accuracy of $\pm10^\circ$. This phenomenon is shown not to be
explained on the basis of traditional notion. A possible explanation is
suggested basing on the hypothesis that there exists a new anisotropic
interaction caused by the cosmological vectorial potential \textbf{A}$_{{\rm
g}}$, a new fundamental constant having, according to the experiments
carried out, the coordinate of right ascension $\alpha \approx  285^\circ$ in the second
equatorial system. This is in agreement with earlier experiments.}

\bigskip
PACS number: 23.40, 12.60i

\pagebreak
\textwidth=155mm

{\large \bf 1. Introduction.}
\vskip10pt
In the Refs. [1-3], periodical variations in the $\beta$-decay rate of $^{{\rm
60}}$Co, $^{{\rm 137}}$Cs, and $^{{\rm 90}}$Sr were
first found. An analysis of 24-hour period of these variations as well as
diurnal rotation of the Earth in different seasons of the year has led to
distinction of some spatial direction characterized by a change in the rate
of the radioactive decay in the vicinity of the point of tangency of the
tangent line to the parallel of latitude when that tangent line has an angle
of $\sim  45^\circ$ with the above mentioned direction. The duration of
uninterrupted measurements in those experiments was no more than two weeks.

In the experiments carried out at JINR (Dubna) from 9.12.1998 till
30.04.1999 with the aim of investigation of variations of the $\beta$-decay rate
of $^{{\rm 137}}$Cs and $^{{\rm 60}}$Co, factors of
experimental instability (for example, the temperature) and their influence
on the final results were taken into account [4]. 24-hour and 27-day periods
of $\beta$-decay rate variations were found. One confirmed also the existence of
the above indicated separate direction in space. The refined spatial
direction in the second equatorial system had the coordinate of right
ascension $\alpha  = 285^\circ$. The declination $\delta$ was not determined in those
experiments.

An analysis of the work [5] in which periodical changes in flows of
particles ($\sim $24 hours) during the $\alpha$ and $\beta$-decay of radioactive
elements were observed, too, is also indicative of the existence of the
above said anisotropic property of the physical space.

The main shortcoming of the works above listed is a high level of
statistical and systematical instabilities and backgrounds masking the
effect. For revealing it one used a special mathematical procedure of
processing the experimental results [2-4]. In addition, the duration of
measurements in some experiments with considerable changes in the $\beta$-decay
rate (the deviation $\approx 7\sigma$ where $\sigma$ is the standard one), was not long ($\sim $
3 days [1]).

In the present paper, the results of new experimental investigations of the
$\beta$-decay of $^{{\rm 137}}$Cs and $^{{\rm 60}}$Co with the
aid of detectors allowing to minimize the statistical and systematical
uncertainties and more clearly reveal the physical phenomenon, are
presented. Briefly outlined are only preliminary results of an experiment
being performed simultaneously in the course of several months at INR RAS
(Troitsk) and at JINR (Dubna).

\bigskip

{\large \bf 2. Measurement Procedure.}
\vskip10pt

The aim of the experiment was to measure, for greater plausibility of its
results, changes in the $\beta$-decay rate simultaneously at different points on
the Earth and uninterruptedly during a long period of time.

The town Dubna and Troitsk are nearly at the same (Moscow's) meridian, the
distance between them is $\sim $ 140km. The experiment lasted from the
middle of February till the middle of May, 2000.

The experimental technique (schematic diagram, electronic equipment,
measurement of backgrounds) at Dubna was the same as in the experiment
described in detail in Ref. [4], and differed from the latter only in that a
Ge(Li)-detector was used. As before (when using a scintillation detector),
the intensities of $\gamma$-transition from an excited level of a daughter nucleus
with an energy of 0.661MeV in the course of the $\beta$-decay of $^{{\rm 1}{\rm
37}}$Cs, was measured. The value of the integral of $\gamma$-quanta counts
entered into the memory of a computer every 10s

As distinguished from the technique of long-term measurements of $\beta$-decay
rate with an scintillation detectors [1-4], the $\gamma$- registration following
the $\beta$-decay of the investigated radioactive nuclei with a Ge(Li)-detectors
made it possible to substantially improve the stability and reliability of
long-term measurements. At Troitsk, a Ge(Li)-detector with a volume of 100
cm$^{{\rm 3}{\rm} {\rm} }$was used for measuring the $\gamma$- spectra
with energies of 1.117MeV and 1.332MeV accompanying the $\beta$-decay
of $^{{\rm 60}}$Co. A radioactive source was placed beyond of the
vacuum volume at a distance of 7mm from the sensitive surface of the
detector. To protect the detector and preamplifier from the possible
influence of alternating high-frequency and magnetic fields, they were
closed by covers from permalloy and electrolytic copper, and a 10cm layer of
lead served as a shielding from the natural radioactive background. The
signal time constant of the input signal was equal to 0.5$\mu$s at a gain factor
of 10-20 which led the influence of amplitude overload of electronic paths
to a minimum.

To record the amplitude spectra, a fast ADC with off-line storage built into
a personal computer (PC), was used. A control program gave the time of
measuring each spectrum (600s), start time, storage instruction, noted the
time of transcription of a next spectrum to the PC memory, zeroed the
off-line storage, and started a new measurement. The program worked in a
cycle so that the information sequentially accumulated in the memory of PC
through a long time. The final processing of information was made off-line
by integration over the spectrum in various intervals of energy from the
first channel to the maximum energy of photopeak (or only the peak itself)
with the resulting formation of a sequence of numbers reflecting the time
dependence of the $\beta$-decay rate. The statistic-average digital load of the
detector was no more than (2$\div$3)$\cdot$10$^{\rm 4}$counts per
second, i.e. corresponded to the optimum working conditions of the
instrumentation. The statistical accuracy obtained at a one point was 0.03\%
for the radioactive decay of $^{{\rm 60}}$Co.

Thus the measurements were made simultaneously by two identical
Ge(Li)-detectors with two independent and different systems of information
storage in natural conditions spaced 140km apart. One detector measured the
decay of $^{{\rm 137}}$Cs, the other did that of $^{{\rm 6}{\rm
0}}$Co.

\bigskip

{\large \bf 3. Results and Discussion.}
\vskip10pt

In Fig.1 and 2 the results of simultaneous measurement of the $\gamma$-counts rate
of at Dubna ($^{{\rm 137}}$Cs) and at Troitsk ($^{{\rm 60}}$Co), respectively, at a period from 16$^{{\rm 24}}$ (Moscow time)
of 15.03.2000 till 10.04.2000 inclusive. To correlate the results of two
experiments, the values of flows from Dubna were averaged over 600s time
intervals, like those at Troitsk, and additionally low-frequency filtration
was made. The results from Troitsk did not processed altogether. As is seen
from Fig.2, the change in the $\gamma$-count rate ranged in that experiment up to
$\sim $0.7\% of the statistical average. At Dubna these changes were no more
than 0.2\%. It should be noted that the flow jump in the vicinity of
2180min (Fig.1) was caused by a technological change of the radioactive
source relative to the Ge(Li)-detector in the process of refilling a Dewar
flask with liquid nitrogen. The difference in the values of $\gamma$-count rate
changes in the experiments at Dubna and Troitsk is probably connected with
dissimilar measuring procedures: in the former case the $\gamma$-quanta were
counted at a fixed energy threshold but at Troitsk one measured the total
amplitude spectra and then determined the total $\gamma$-counts by integration over
the spectrum. The difference of deviations can be explained also by that
magnetic moments of the nuclei of $^{{\rm 60}}$Co and $^{{\rm 137}}$Cs. However that difference can be possibly associated with the
different latitude position of the experimental setups.

The observed regular structure in the time dependence of the $\beta$-decay rate
for nuclei $^{{\rm 137}}$Cs and $^{{\rm 60}}$Co can be, in
general, explained by the following reasons:

\noindent
\item{a) temporal instabilities of the electronic recording paths;}

\noindent
\item{b) outside influences and those connected with the human activity;}

\noindent
\item{c) unknown physical processes in the Ge(Li)-detectors themselves in the course
of long-term measurements;}

\noindent
\item{d) a ``cosmological'' factor acting on the process of the $\beta$-decay of nuclei.}

Consider each reason separately for the setup of Troitsk (that of Dubna, as
was mentioned above, is considered in detail in Refs [3,4]).

\item{a) The structure of changes in the count rate is such that it cannot be
explained by the slow variations of the supply-line voltage (220V). Besides,
the low-voltage supply of the electronic circuits was stabilized with an
accuracy of 3\% and did not vary when the supply-line voltage was changed
within 15\%. As is known, the spectrometric characteristics of
Ge(Li)-detectors practically do not depend on insignificant variations of
high-voltage supply.}

\item{b) As for the structural changes in the $\beta$-decay count rate due to variations of
some external influences, daily variations of intensity of cosmic radiation,
changes in the room temperature, etc, we can say that the detectors
themselves as well as the channels of electronic paths were carefully
protected against variable, alternating, and leakage electromagnetic and
high-frequency fields. The natural background of the Ge(Li)-detector was as
little as about 0.1\% of the $\beta$-decay count rate of the radioactive sources
investigated. That is, only a periodical 7-8 times increase of the
background could explain the structure observed in the $\beta$-decay count rate.
But the count rate of the background was constant in the limits of the
statistical accuracy. To evaluate experimentally a possible influence of
count overloads (idle time) of the electronic paths when measuring the
background, one simultaneously fed to the input of the preamplifier a signal
from a generator with an amplitude equal to that of the photopeak but with
frequency 50-100 times more than under the operating conditions. For such
spectra, there were no peculiarities in the time dependence of the count
rate.}

\item{c) It is possible that in the material itself of the Ge(Li)-detector some yet
unknown physical processes take place during the long-term exposure to
radiation which lead to accumulation of charge in ``internal'' capacities,
then to a break-down and relaxation of charge. This would correspond to the
observable structure form in the time dependence if the ``time constant''
were close to 24 hours. In such a case the amplitude spectrum of $\gamma$-quanta
would be perturbed, too, in those intervals of time but this was not
observed in the experiments. As to an influence of capacitive coupling in
the amplifying section itself, control measurements were carried out with
the use of non-capacitive current amplifiers, and a similar structure with
the same value was found in the time dependencies obtained.}

\item{d) Finally consider a possible influence of the cosmological factor on nuclear
processes on Earth [1-4,6,7,9]. As was said [6,7,9], it can be associated
with a new suggested interaction of objects in nature caused by the
existence of the cosmological vectorial potential \textbf{A}$_{{\rm g}}$, a
new fundamental vectorial constant entering into the definition of byuons,
discrete objects. According to the hypothesis advanced in Refs. [6-9], in
the process of minimization of byuons' potential energy in the
one-dimensional space formed by them, the three-dimensional space R$_{{\rm
3}}$ and the world of elementary particles appear together with all their
quantum numbers and their quantum-mechanical behavior in R$_{{\rm 3}}$.}

In the model considered, the masses of particles are proportional to the
modulus of the summary potential \textbf{A}$_{{\rm \Sigma}}$. It comprises
\textbf{A}$_{{\rm g}}$ and the vectorial potentials of magnetic sources as
of natural (from the Earth, the Sun, etc.) so of artificial origin (for
example, the vectorial potentials \textbf{A} of magnetic fields of plasma
generators, solenoids etc.). The magnitude $\vert $\textbf{A}$_{{\rm
\Sigma}}\vert $ is always lesser than $\vert $\textbf{A}$_{{\rm g}}\vert $ $\approx$
1.95$\cdot$10$^{{\rm 11}}$Gs$\cdot$cm [6-9]. Practically the vector
\textbf{A}$_{{\rm \Sigma}}$ is collinear to \textbf{A}$_{{\rm g}}$ due to the
great value of the latter.

In the model [6,7], the process of formation of physical space and charge
numbers of elementary particles is investigated. Therefore the values of
potentials acquire a physical sense (as distinct from calibration theories,
for example, the classical and quantum field theories) which is in tune with
the known and experimentally tested Aharonov-Bohm effect [10-13] that is a
special case of quantum properties of space described in Refs. [6,7].

According to the terrestrial experiments with high-current magnets
[6,7,14-16], with a gravimeter and attached magnet [6,7,17], as well as in
accordance with investigations of changes in $\beta$-decay rate of the radioactive
elements under action of the new force [1-4,6,7,9] and astrophysical
observations [6,7,18,19], the direction of the \textbf{A}$_{{\rm g}}$ had the
following rough coordinates: right ascension $\alpha\approx 270^\circ$, declination $\delta \approx 34^\circ$.

More accurate experiments with a stationary [20] and pulsed [21] plasma
generators placed on rotatable bases for scanning the celestial sphere have
given the coordinates $\alpha = 293^\circ \pm 10^\circ, \delta = 36^\circ \pm 10^\circ$.

The experiments listed have shown that if the vectorial potential of some
current system is in opposition to the vector \textbf{A}$_{{\rm \Sigma}}$ then
the new force pushes any substance out of the region with weakened $\vert
\textbf{A}_{{\rm \Sigma}}\vert$ mainly to the side of \textbf{A}$_{{\rm g}}$.

The new force predicted in Refs. [6,7,14-16] is of complex nonlinear and
nonlocal character and can be represented in the form of some series in
$\Delta$A$_{{\rm \Sigma}}$,\textit{} a difference between changes in A$_{{\rm \Sigma}}$ at the points
of location of sensor and test body. The expansion of this series in terms
of $\Delta$A$_{{\rm \Sigma}}$ gives as a first approximation

\begin{equation}
\label{eq1}
F\sim \Delta A_{\Sigma}  {\frac{{\partial {\kern 1pt} \Delta A_{\Sigma}
}}{{\partial {\kern 1pt} x}}}
\end{equation}

\noindent
where x is the spatial coordinate in R$_{{\rm 3}}$.

The scaling estimation of magnitudes of potentials of Earth's (A$_{{\rm
E}}$) and Sun's (A$_{{\rm \odot} }$) magnetic fields in the range of the Earth's
orbit show that they are equal to $\sim $ 10$^{{\rm 8}}$Gs$\cdot$cm and $\sim $
5$\cdot$10$^{{\rm 8}}$Gs$\cdot$cm, respectively. Thus when rotating a setup for
measuring the count rate of the $\beta$-decay, one can observe variations of the
modulus of \textbf{A}$_{{\rm \Sigma}}$ of $\sim \vert $\textbf{A}$_{{\rm E}}$
+ \textbf{A}$_{{\rm \odot} }\vert $ / $\vert $\textbf{A}$_{{\rm \Sigma}}\vert $ $\approx$
10$^{{\rm -3}}$ around the Earth and the Sun.

As the first approximation of the new force in terms of $\Delta$A$_{{\rm \Sigma}}$
contains not only $\Delta$A$_{{\rm \Sigma}}$ but ${\frac{{\partial {\kern 1pt} \Delta
A_{\Sigma} } }{{\partial {\kern 1pt} x}}}$, too, one can suggest that the
change $\Delta$W in the probability of the $\beta$-decay should be proportional to
$\Delta A_{\Sigma}  {\frac{{\partial {\kern 1pt} \Delta A_{\Sigma}
}}{{\partial {\kern 1pt} x}}}$ [9] but not only to $\Delta$A$_{{\rm \Sigma}{\rm} }$as
was meant$_{{\rm} }$ in Refs. [6,7]. Hence the new force can influence on
the decay of the neutron.

Let us explain the aforesaid.

As is known [22], the neutron has a magnetic moment M$_{{\rm n}}\approx10^{{\rm 23}}$erg/Gs.
In this connection and according to scaling
estimations it is likely that there exists an enormous value of
${\frac{{\partial {\kern 1pt} \Delta A_{\Sigma} } }{{\partial {\kern 1pt}
x}}}\approx 10^{{\rm 16}}$Gs in the vicinity of the neutron but therewith
the magnitude $\Delta$A$_{{\rm \Sigma}}$ from the magnetic field is not high and is
equal to 10$^{{\rm 3}}$Gs$\cdot$cm. As during the rotation of radioactive sources
together with the Earth around its axis and the Earth's motion around the
Sun the $\Delta$A$_{{\rm \Sigma}}$ can vary within five order of magnitude at the places
of their location by the action of potentials A$_{{\rm E}}$ and A$_{{\rm \odot
}}$, so the joint action of two factors of $\Delta$A$_{{\rm \Sigma}}$ from the Earth and
the Sun as well as ${\frac{{\partial {\kern 1pt} \Delta A_{\Sigma}
}}{{\partial {\kern 1pt} x}}}$ from the neutron's magnetic field can create
a value of the new force sufficient to influence on the period of decay of
the neutron.

In Fig.3 and 4, the small circles denote places of observation of maximum
flows of $\gamma$-quanta at the $\beta$-decay of $^{{\rm 60}}$Co at Troitsk as well
as the maximum and minimum flows for the decay of $^{{\rm 137}}$Cs at Dubna, respectively.
The numbered maxima and minima of flows of $\gamma$-quanta in Figs.1 and 2 correspond to number of arrows in Figs.3,4 drawn
from place of observation of an extremum tangentially to the parallels of
latitude (along which the vectorial potential of the magnetic field of the
Earth's dipole is directed)

As is seen from Figs.3,4 the total set of numbered arrows can be clearly
divided, by their directions, into three subsets with an accuracy of $\pm10^\circ$.

In Fig.4 we have the subsets D$_{{\rm 1}}$ (7, 9, 11, 13, 17, 19, 21, 23,
25, 33, 39a, 41, 43, 3,6, 10, 16a, 18, 20, 22, 24a, 42), D$_{{\rm 2}}$ (2b,
4, 14, 34, 1, 42, 44), D$_{{\rm 3}}$ (2a, 12, 16b, 24b, 39b, 5). In Fig.3
these are: T$_{{\rm 1}}$ (1, 2, 3, 4, 5, 6, 7, 10, 11, 15, 17, 18, 21),
T$_{{\rm 2}}$ (8, 9, 19, 20), T$_{{\rm 3}}$ (13, 22, 12, 16). The subset
D$_{{\rm 1}}$ is seen to be in correspondence with T$_{{\rm 1}}$, D$_{{\rm
2}}$ is with T$_{{\rm 2}}$, and D$_{{\rm 3}}$ is with T$_{{\rm 3}}$. Thus we
see, in the same time interval and at the different experimental setups
being 140km apart, a similar pattern of behaviour of the maxima and minima
of the $\beta$-decay rate for the different radioactive elements $^{{\rm 1}{\rm
37}}$Cs and $^{{\rm 60}}$Co.

The obtained subsets D$_{{\rm 1}}$,T$_{{\rm 1}}$ and D$_{{\rm 3}}$,T$_{{\rm
3}{\rm} }$ clearly correspond (with an accuracy of $\pm$10$^\circ$) to the directions
of action of the new force fixed during scanning the celestial sphere by the
stationary [20] and pulsed [21] plasma generators. These directions were
found from dependences of heat release in the jet of the plasma generators
on the angle.

The experiments with the pulsed plasma generator [21] have shown that at
each point of space there exists a cone of directions of action of the new
force with an opening $\sim $100$^\circ$. Therewith the axial line of the cone is
directed along the vector \textbf{A}$_{{\rm g}}$ to which, according to the
results obtained (see Figs.3,4), an angle $\alpha \approx 285^\circ$ corresponds that fall
into the range of magnitudes of the coordinate $\alpha$ for the vector
\textbf{A}$_{{\rm g}}$ (on evidence of Refs. [21]) and is near to the
earlier results [6,7,14-16]. Thus the sets of arrows D$_{{\rm 1}}$T$_{{\rm
1}}$ and D$_{{\rm 3}}$T$_{{\rm 3}}$ coincide with the directions of the cone
generator to the precision indicated.

The emergence of the set D$_{{\rm 2}}$, T$_{{\rm 2}}$ is easily explicable
on the basis of a hypothesis of equiprobable distribution of directions of
neutron magnetic moments in the nuclei and total substance of radioactive
source. That is, in any source always there are neutrons with the magnetic
moments being perpendicular to the vector \textbf{A}$_{{\rm g}}$ and hence
the lines of the vectorial potential of magnetic field of the neutron will
be always directed, in some region of space, under an efficient angle of
$\sim 140^\circ$ to \textbf{A}$_{{\rm g}}$ [20,21] corresponding to the maximum
value of ${\frac{{\partial {\kern 1pt} \Delta A_{\Sigma} } }{{\partial
{\kern 1pt} x}}}$ and hence of the new force.

The authors are grateful to the academicians of RAS Belyaev S.T. and Matveev
V.A., to the corresponding members of RAS Lobashov V.M., Barashenkov V.S.,
and to all participants of seminars at INR (Troitsk) for useful advices and
fruitful discussions of the results obtained, as well as to Morozov E.P.,
Kazinova L.I., and Baurov A.Yu. for the help in preparing the text of the
paper.

\begin{figure}[h]
\centerline{\psfig{figure=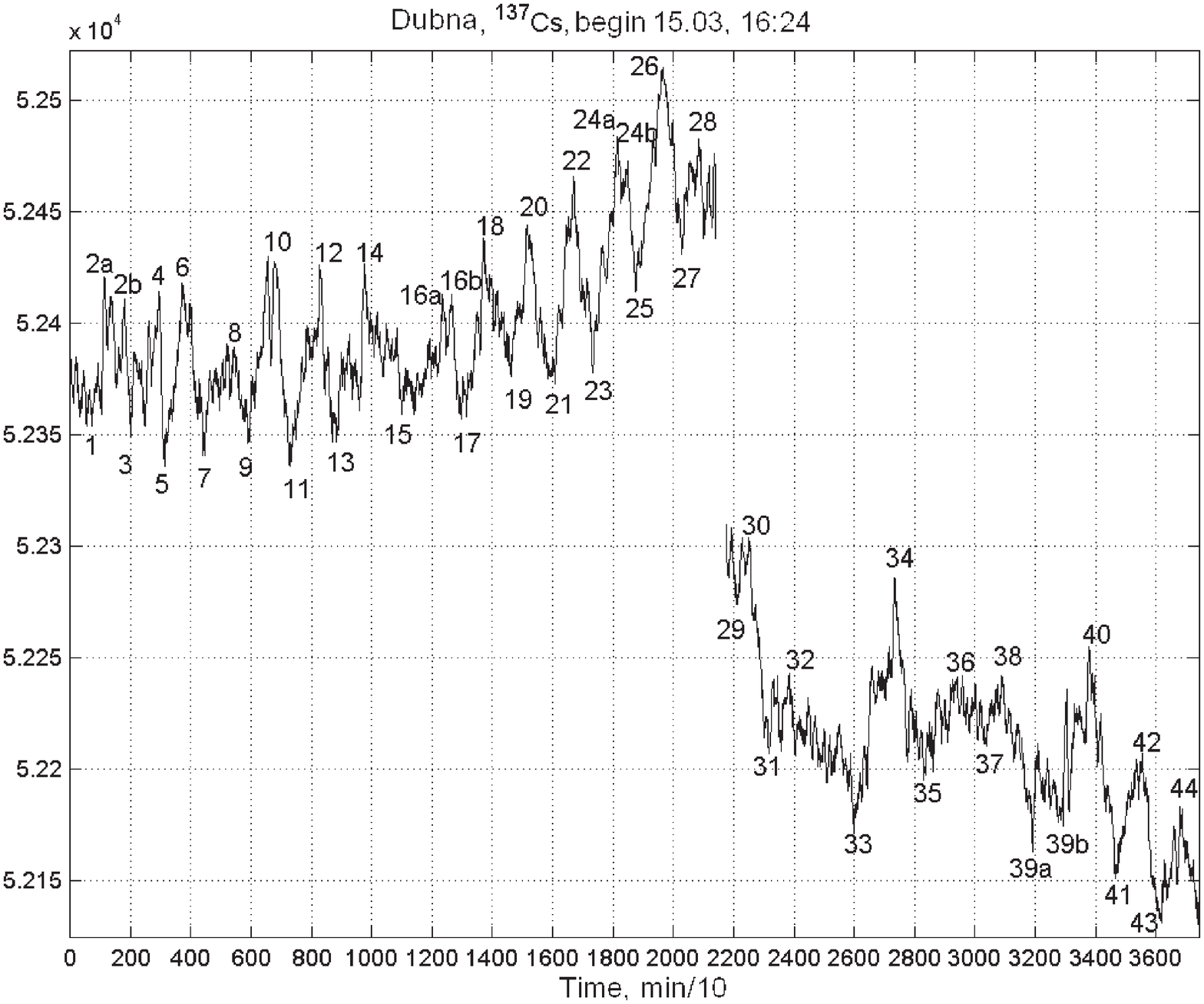,angle=0,height=100mm,width=125mm}}
\end{figure}

Fig.1 The variation of the flow of $\gamma$-quanta accompanying the $\beta$-decay of $^{{\rm 137}}$Cs,
 with the time (JINR, Dubna).

\begin{figure}
\centerline{\psfig{figure=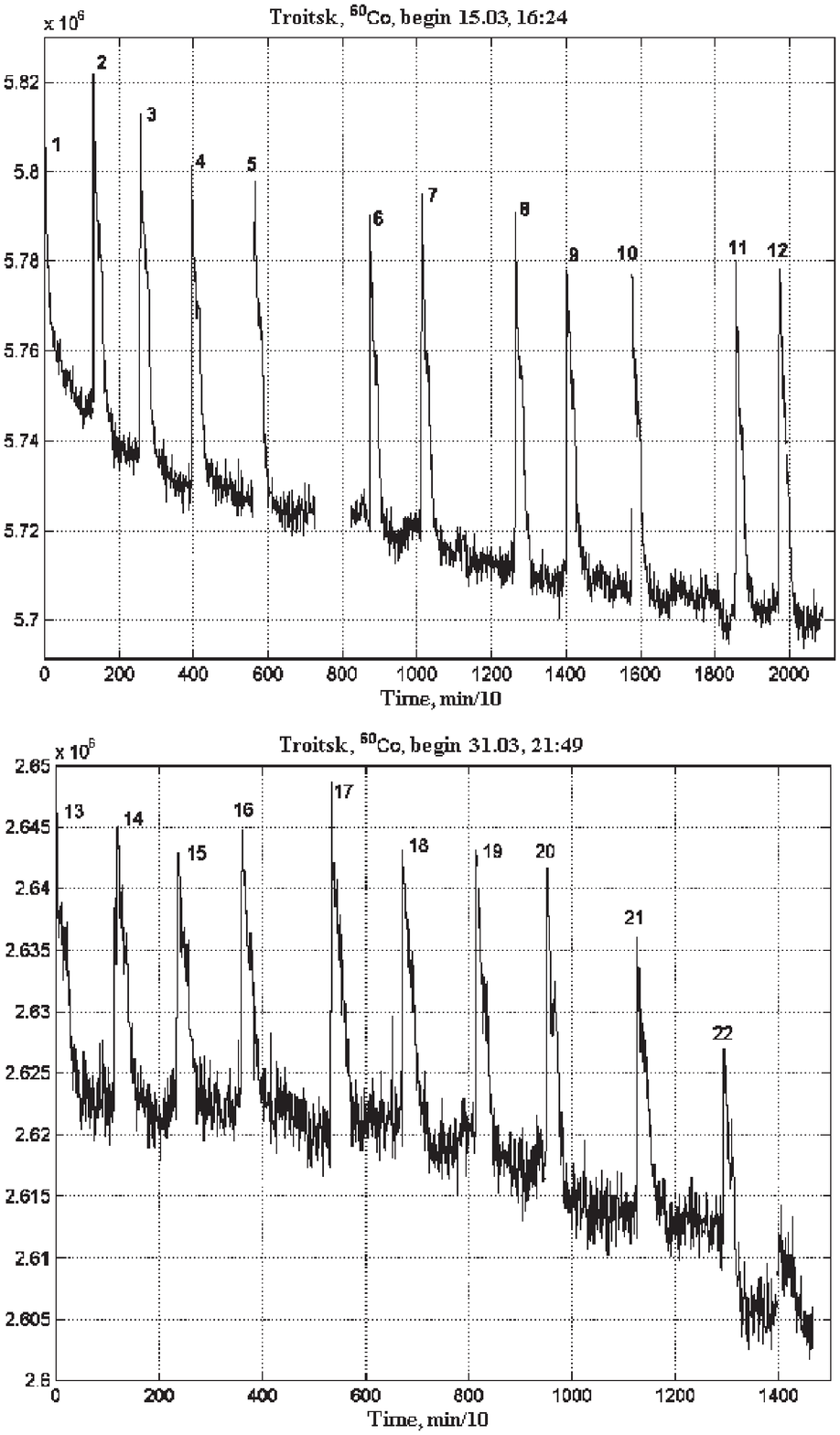,angle=0,height=170mm,width=110mm}}
Fig.2. The variation of the flow of $\gamma$-quanta accompanying the $\beta$-decay of
$^{{\rm 60}}$Co, with the time (INR, Troitsk).
\end{figure}

\pagebreak

\begin{figure}[ht]
\centerline{\psfig{figure=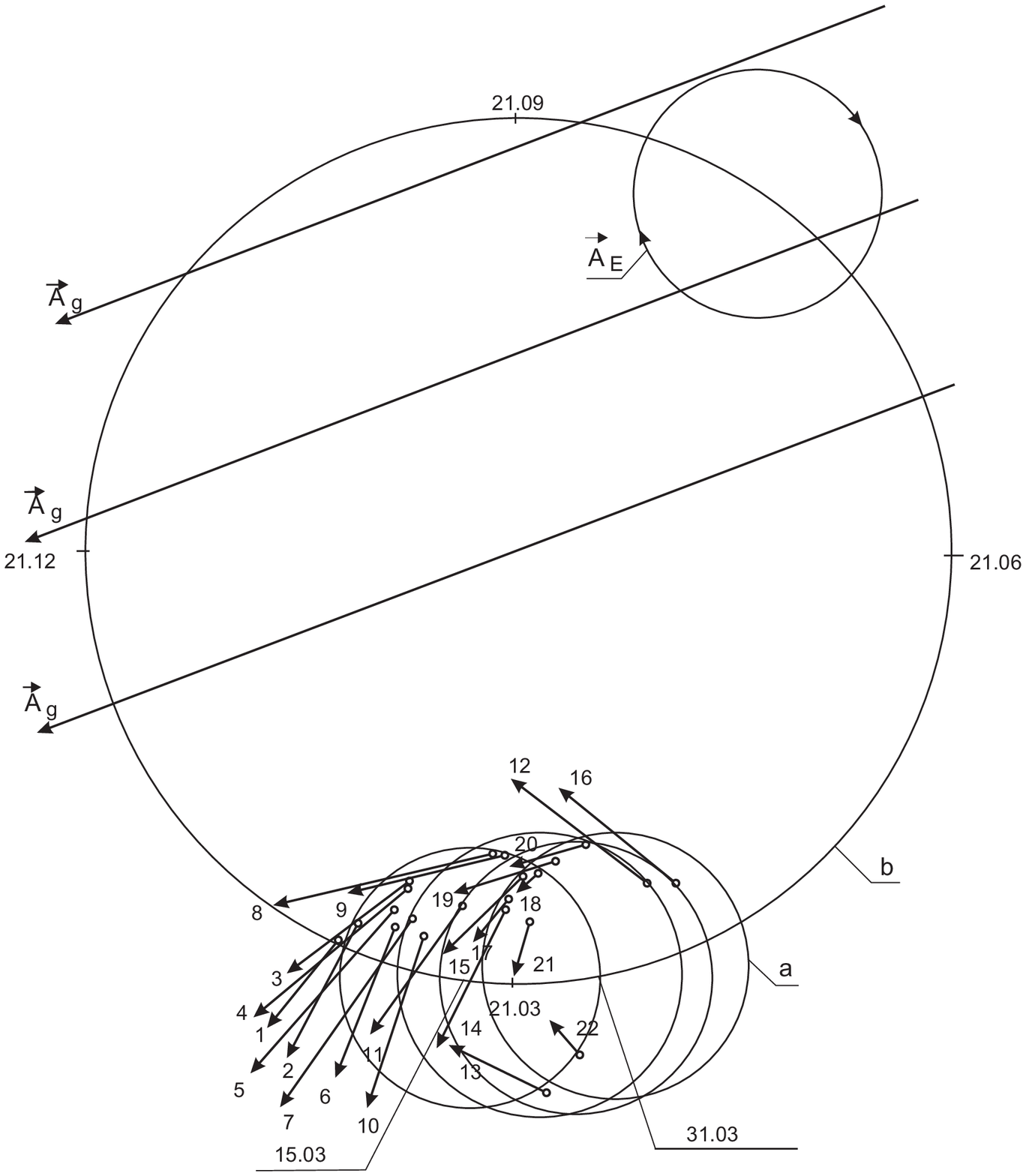,angle=0,height=100mm,width=90mm}}
\end{figure}

Fig.3. The spatial positions of sites where the clearly expressed extrema in
the magnitude of the flow of $\gamma$-quanta in the experiment with the $\beta$-decay of
$^{{\rm 60}}$Co, were observed (see Fig.2).

$\leftarrow${\hskip-2pt}$\bullet$ - the site of the maximum flow of $\gamma$-quanta with the indication of the
direction of action of the new force drawn along the tangent line to the
parallel of latitude;

a -- the trajectory of motion of the radioactive source rotating together
with the Earth;

b -- the trajectory of motion of the Earth and the radioactive source around
the Sun;

21.03 etc. -- the point of the vernal equinox and other characteristic points
of the trajectory ``b'';

$\vec {\rm A}_{\rm E}$ -- the direction of the vectorial potential of the magnetic
field of the Earth's dipole;

$\vec {\rm A}_{\rm g}$ -- the direction of the cosmological vectorial potential.

\pagebreak

\begin{figure}[ht]
\centerline{\psfig{figure=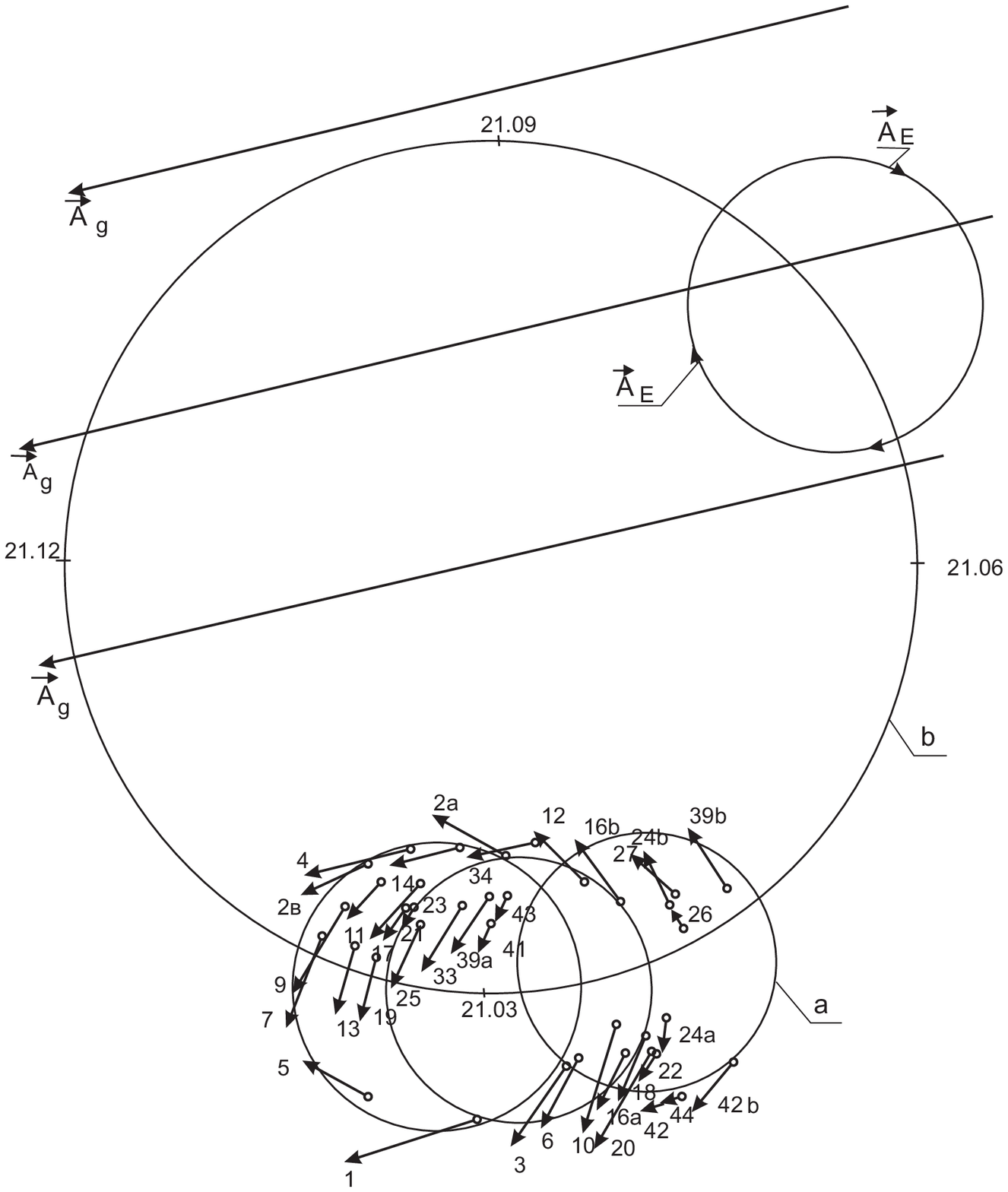,angle=0,height=95mm,width=80mm}}
\end{figure}

Fig.4. The same as in Fig.3 but for minimum and maximum flows of $\gamma$-quanta
during the $\beta$-decay of $^{\rm 137}$Cs (see Fig.1).

\end{document}